# Water Production Rates from SOHO/SWAN Observations of Comets C/2017 K2 (PanSTARRS) and C/2022 E3 (ZTF)

M.R. Combi[1*], T. Mäkinen[2], J.-L. Bertaux[3], E. Quémerais[3], and S. Ferron[4]

[1]Dept. of Climate and Space Sciences and Engineering
University of Michigan
2455 Hayward Street
Ann Arbor, MI 48109-2143
United States of America
*Corresponding author: mcombi@umich.edu

[2]Finnish Meteorological Institute, Box 503
SF-00101 Helsinki, FINLAND

[3]LATMOS/IPSL
Université de Versailles Saint-Quentin
11, Boulevard d'Alembert, 78280, Guyancourt, FRANCE

[4]ACRI-st, Sophia-Antipolis, FRANCE

**ABSTRACT**

In 2022 and 2023 the hydrogen comae of two long period comets, C/2017 K2 (PanSTARRS) and C/2022 E3 (ZTF), were observed with the Solar Wind ANisotropies (SWAN) all-sky hydrogen Lyman-alpha camera on the SOlar and Heliosphere Observer (SOHO) satellite. SWAN obtains nearly daily full-sky images of the hydrogen Lyman-alpha distribution of the interstellar hydrogen as it passes through the solar system yielding information about the solar wind and solar ultraviolet fluxes that eat away at it by ionization and charge exchange. The hydrogen comae of comets, when of sufficient brightness, are also observed. Water production rates have been calculated over time for each of these comets, covering about 6 months mostly of the post-perihelion period of C/2017 K2 (PanSTARRS) and about 3 months around perihelion of C/2022 E3 (ZTF).

**1. INTRODUCTION**

Two long period comets were observed by the Solar Wind Anisotropies (SWAN) all-sky hydrogen Lyman-$\alpha$ (H Ly$\alpha$) camera on the Solar and Heliosphere Observer (SOHO) satellite during 2022 and 2023: C/2017 K2 (PanSTARRS) and C/2022 E3 (ZTF). The principal investigation of SWAN is observing all-sky images at H Ly$\alpha$ to measure the changing spatial structure of interstellar hydrogen streaming through the solar system. This provides a global changing picture of the solar wind as it depletes interstellar hydrogen while streaming through the solar system (Bertaux et al. 1995). It also provides an excellent vehicle for observing the H Ly$\alpha$ emission from the extended hydrogen comae of comets. SOHO and SWAN have been operating since being inserted into a halo orbit around the L1 Sun-Earth Lagrange point after its launch in December 1995. From here the suite of SOHO's instruments detect the solar wind and its composition shortly before striking the Earth and image the Sun and its corona at a variety of wavelengths.

Comets are surrounded by a huge atmosphere, or coma, of atomic hydrogen that is produced mainly by the photodissociation first of water, typically the most abundance volatile species, producing OH + H and in a second step of dissociating OH producing O + H. Because of the excess energy resulting from the UV photons required for the dissociation, H atoms are ejected at speeds with major components of 8 and 20 km s$^{-1}$ (Combi et al. 2004). Heavy components such as O atoms and OH radicals are much slower. In moderate to very active comets the fluorescence scattering of solar H Ly$\alpha$ photons produces a large visible hydrogen coma easily detectable by the SWAN instrument.

Since SOHO's launch in December 1995, SWAN has observed over 70 comets totaling over 90 individual comet apparitions with several short period comets having been observed several times each (Combi et al. 2019; Combi 2022). Given that two H atoms are produced per water molecule sublimated from the surface of the comet nucleus, water production rates can be calculated from each H Ly$\alpha$ image of the comet observed by SWAN using the methods described in detail by Mäkinen and Combi (2005). The method uses the detailed physics of water outflow, photodissocation, the exothermal velocities on H atoms and their subsequent partial thermalization on their way through the heavy molecule inner coma. See the physical description by Combi et al. (2004) and the references therein for the details and demonstrated observational constraints.

A summary of the SWAN observations of comets C/2017 K2 (PanSTARRS) and C/2022 E3 (ZTF) and their orbital parameters are given in Table 1. In the remainder of this paper, the production rate uncertainties, dQ, are 1-sigma values from the fit of the variation of the model spatial distributions as scaled to the observation as well as the fitted interplanetary hydrogen background and noise in the data. Systematic uncertainties that result from the solar Lyman-alpha flux as obtained from LASP (http://lasp.colorado.edu/lisird), the calibration of the SWAN itself, as well as those of the model parameters, combined with faint field stars not able to be explicitly accounted total on the order of 30%.

## 2. C/2017 K2 (PanSTARRS)

Comet C/2017 K2 (PanSTARRS), referred to hereafter as comet K2, is a dynamically new Oort Cloud comet (OCC) with an original semi-major axis of 28000 au according to the definitions by A'Hearn et al. (1995) where dynamically new comets have a semi-major axis > 20000 au. It was discovered when it was 16 au from the Sun, well beyond the orbit of Saturn, with the PanSTARRS survey telescope by Wainscoat et al. (2017). Hui et al. (2017) found pre-discovery images dating back to at least May 2013 when it was at a heliocentric distance of 23.7 au. Observations of CO in the coma of comet K2 were reported by Yang et al. (2021) using the James Clerk Maxwell telescope when it was at a heliocentric distance of 6.72 au, well outside the orbit of Jupiter and well before atomic hydrogen was detected by SWAN, yielding a CO production rate of (1.6 ± 0.5) x $10^{27}$ $s^{-1}$. Cambianica et al. (2023) have reported observations of the forbidden green and red emissions of atomic oxygen in comet K2 with ratios that suggest the presence of significant amounts of $CO_2$ photodissociation at a heliocentric distance of 2.8 au and that activity might have been driven by $CO_2$ sublimation well outside the sublimation temperature of $H_2O$.

The H coma of comet K2 was observed by the SWAN instrument on SOHO from 28 October 2022 to 25 April 2023 with only six usable images before perihelion but 83 images covering post-perihelion heliocentric distances continuously from 1.8 to 2.4 au. Water production rates were calculated using the method described in detail by Mäkinen and Combi (2005) from each of the 89 images. These results given in Table 2 and are shown plotted as a function of time from perihelion in days. The post-perihelion water production varied with a power-law exponent in the heliocentric distance of -3.2, a fairly normal variation for an OCC. The pre-perihelion had too few observations to obtain a good power-law fit, but it appears qualitatively similar. The water production rates, their 1σ uncertainties as well as observational geometry parameters are given in Table 2 and are plotted as a function of time in days from perihelion in Figure 1 and compared with the few of the available contemporaneous observations. There was one fairly well-defined factor of ~2 outburst that lasted 15 days from beginning to end and peaked at 68 days after perihelion on 26 February 2023. The fact that K2 is a dynamically new comet probably explains the large activity and CO production rate far from the Sun pre-perihelion but rather flat activity around and the typical drop in production rate well after perihelion (Combi et al. 2019).

As of this writing most of the other published (Kwon et al. 2024; Ejeta et al. 2025) and preprint values of water production rates in K2 were taken pre-perihelion before most of our SWAN observations. The short-term variation of our SWAN-determined water production rates is quite wide. The pre-perihelion range within 50 days of perihelion is similar to that post-perihelion. The more distant pre-perihelion water production rates from infrared observations by Ejeta et al. (2025) are reasonably consistent given the wide range of post-perihelion SWAN results. Otherwise, hopefully publication of the SWAN result will serve as some guide to future publications of observations already taken. Figure 1 also shows one value from a et al. (2022) derived from ground-based OH observations. Given the generally large scatter in the SWAN results owing to the fact that while the production rates were rather moderate, K2's heliocentric and geocentric distances are rather large and so the intrinsic Lyman-alpha brightness of the coma was not large and so the many faint background stars which could not be identified and accounted for contributed to the wide scatter. In any case, given the wide scatter the result from the one OH measurement is not seen as a serious inconsistency with the SWAN results.

**3. C/2022 E3 (ZTF)**

Comet C/2022 E3 (ZTF), referred to as comet E3 hereafter, is an older Young Long-Period OCC, having an original semi-major axis of only 1300 au, on the shorter end of the Young Long-Period comets (a > 500 au) and meaning that it has definitely passed through the inner solar system some 47000 years in the past. Gravitational interactions throughout the perihelion passage indicate that it is now likely on a hyperbolic orbit and so it may never return. It was discovered by Bolin et al. (2022) on 2 March 2022 using the Zwicky Transient Facility 1.2-m Palomar telescope.

Table 3 shows the water production rates determined from the SOHO/SWAN images of the H Lyα coma. E3 has an unusual variation in water production as a function of time/heliocentric distance, indicating a likely very strong seasonal effect. The variation over its orbit is reminiscent of comet C/2009 P1 (Garradd) which was a truly dynamically new comet and was found to have been shedding a rather strong emission of icy grains on the entire pre-perihelion leg (Combi et al 2013), followed by a more typical drop in production rate with distance after perihelion. Implications of the icy grains were also supported in a later more extensive comparison of various observations by Feaga et al. (2014). For E3 this behavior could indicate a seasonal effect where the pole (or face) of the nucleus primarily facing the Sun pre-perihelion has a strong emission of icy grains but that facing the Sun post-perihelion does not. Clearly there are also Jupiter Family Comets that have activity dominated by the release of icy grains that provide an extended source of coma gas (Combi et al. 2020b).

Schleicher et al. (2023) reported the pre-perihelion production rates were a factor of two larger than the post-perihelion values. This asymmetry is similar in magnitude to the SWAN results with the average pre-perihelion values of about 6 x $10^{28}$ s$^{-1}$ and the post-perihelion value of about 3 x $10^{28}$ s$^{-1}$. However, the largest value according to Schleicher (private communication) of 4.8 x $10^{28}$ s$^{-1}$ actually occurred 2 weeks after perihelion on 28 January when the values found in the SWAN results were only about 3/4 of this value. The larger set of SWAN values shows that an abrupt change in production rate occurs at about 15 days before perihelion which could indicate the approximate time of the main seasonal change (equinox on the nucleus) or the complete sublimation of icy grains accumulated during the more distant pre-perihelion part of its orbit as appeared to be the case for comet C/2009 P1 (Garradd). Finally, Schleicher et al. also reported a rotation period of K2 of 8.7 +/- 0.1 hours whereas Manzini et al. (2023) reported a similar rotation period of 8.49 ± 0.12 hours.

It is worthwhile to note that the visual magnitude (http://www.aerith.net/comet/catalog/2022E3/2022E3.html), which is determined mainly by the dust production, shows a very different distribution that is more symmetrical than water about a peak value about a month after perihelion, but away from the peak the dust production is also generally larger before perihelion than after. While not typical in comets, the similarity with comet C/2009 P1 (Garradd) having large pre-perihelion water production compared with post-perihelion combined with the generally typical rather symmetric variation of visual magnitude (dust) is perhaps noteworthy.

Biver et al. (2024) observed comet E3 with the IRAM 30-m telescope from 3 to 8 February 2023 and detected the 183.3 GHz water line. They also reported observations with the Nançay Radio Telescope and report OH production rates increasing from 3.3 to 8.1 x $10^{28}$ $s^{-1}$ between 17 October and 23 December 2022. After an instrumentation failure they observed again from 17 to 24 February 2023 and only obtained an upper limit value of 6 x $10^{28}$ $s^{-1}$. The water production rate determined by Biver et al. is shown as the filled circle in Figure 2 as is one value from Jehin et al. (2022b) shown as a filled triangle which unlike the values from Schleicher (private communication) and Biver et al. (2024) is much smaller than the SWAN results at the same time. There is no explanation at this time for these larger than normal differences. Biver et al. (2024) also observed individual emissions of HCN, CS, $H_2S$, $H_2CO$, $CH_3CN$, $NH_2CHO$, HNCO, HCOOH, $CH_3CHO$ and CO. From the HCN observations they detected a sunward biased coma production with a dayside expansion velocity of 0.76 km $s^{-1}$ and a nightside expansion of 0.52 km $s^{-1}$ with roughly ¾ of the production on the dayside. The kinematics of the $CH_3OH$ emission were consistent with the HCN result.

## 4. Summary

We describe herein the results of the analysis of the observations of the hydrogen Lyman-alpha comae of two long-period comets C/2017 K2 (PanSTARRS) and C/2022 E3 (ZTF) observed during the 2022-2023 by the SWAN all-sky camera on the SOHO spacecraft. SOHO/SWAN provides an extended time period of water activity in comets and the levels of the water production rates given the uncertainties and scatter in the results are fairly consistent with the other cited published results from OH and $H_2O$ observations. C/2017 K2 (PanSTARRS) is a Very Young and nearly classifiable as a Dynamically new comet. Because of the location of the comet on the detector there were only 6 useful observations pre-perihelion but they were generally consistent with the 77 observations covering 4 months of the post-perihelion orbit that indicated a variation with heliocentric distance with an exponent of -3.2, fairly nominal for the range of heliocentric distances. Generally this kind of behavior is similar to many dynamically new comets (Combi et al. 2019).

Comet C/2022 E3 (ZTF), on the other hand, is a Very Old Long Period comet having been through the planetary region of the solar system 47000 years ago. Its variation with heliocentric distance was less normal for the range of heliocentric distances covered, not varying anything like a typically used power-law. That said, if a power-law is fitted to the results, the pre-perihelion exponent is -1.3 and the post-perihelion exponent is -0.4. Perhaps if SWAN could have detected the comet to much larger heliocentric distances a more typical variation would appear. It demonstrated a pre- to post-perihelion asymmetry, being much larger before perihelion by roughly a factor of 2. However, it is also noted that the asymmetry in the visual magnitude, dominated by dust production, is quite different with a variation that was more symmetric about a peak that happened about a month after perihelion but still with a larger value pre-perihelion away from the peak. A similar difference between water production and visual magnitude asymmetries about perihelion was also found in dynamically new comet C/2009 P1 (Garradd) where more extensive available observations also showed the larger pre-perihelion water production rates were caused by a large extended source of the icy grains released pre-perihelion but had sublimated away by the time of perihelion. There were also a number of ups and downs in activity throughout with peaks at -55, -30, -5, +8, +20, and +31 days relative to perihelion or roughly every 15 to 25 days. These cannot be related to the ~8-hour rotation period but must deal with some dynamical issue on the nucleus like outburst emissions of icy particles.

**Acknowledgements:** SOHO is an international mission between ESA and NASA. M. Combi acknowledges support from NASA grant 80NSSC23K0030 from the Solar System Observations Program. T.T. Mäkinen was supported by the Finnish Meteorological Institute (FMI). J.-L. Bertaux and E. Quémerais acknowledge support from CNRS and CNES. We obtained orbital elements from the JPL Horizons web site (http://ssd.jpl.nasa.gov/horizons.cgi). For classification of the dynamical ages of the comets we used the Minor Planets Center web site https://www.minorplanetcenter.net/db_search tool. The composite solar Lyα data were taken from the University of Colorado LASP web site at the (https://lasp.colorado.edu/lisird/data). We acknowledge the personnel who have been keeping SOHO and SWAN operational for almost 29 years.

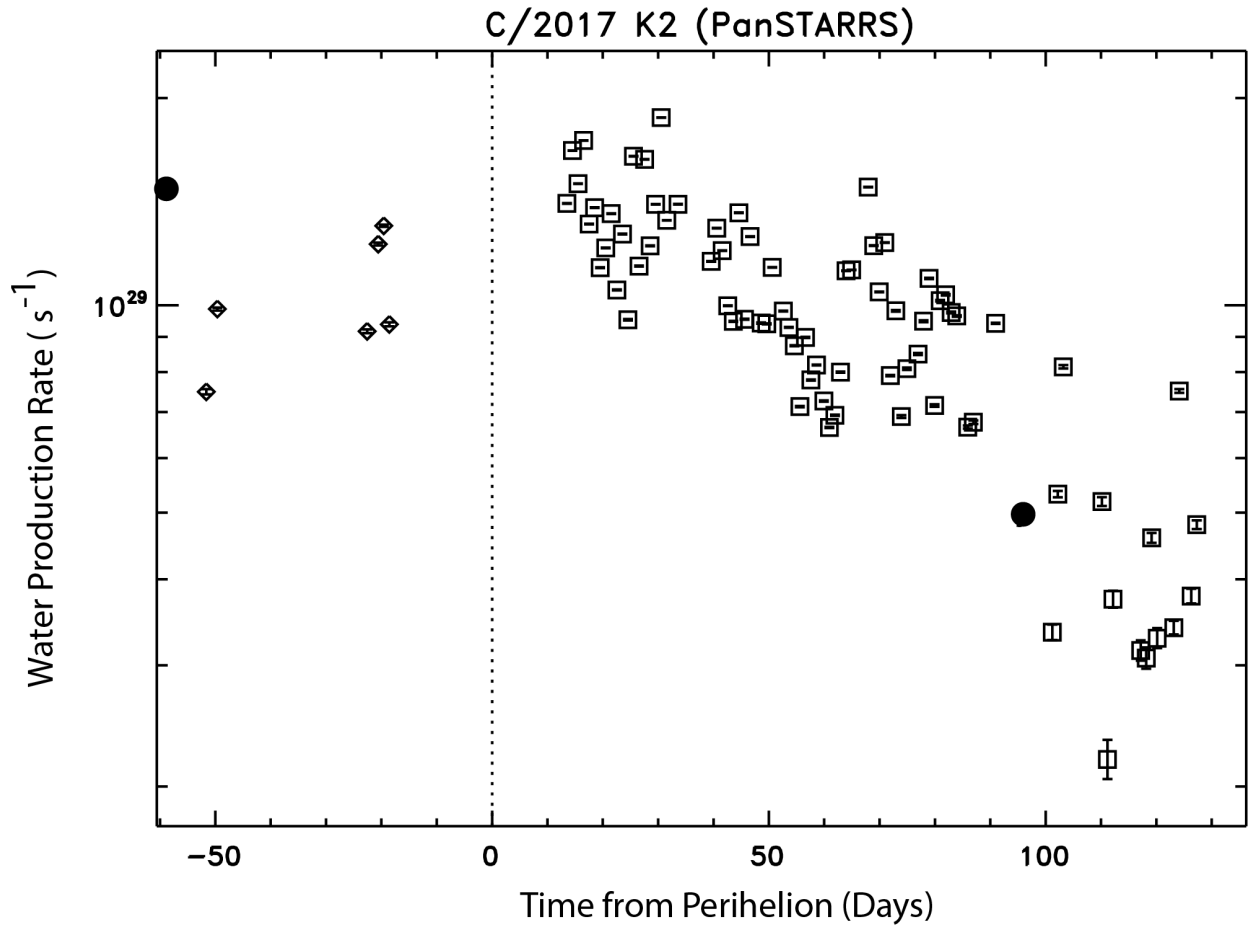

Figure 1. Water production rate of comet C/2017 K2 (PanSTARRS) as a function of time from perihelion. The open symbols give the water production rate in $s^{-1}$ from single images, diamonds pre-perihelion and squares post-perihelion. The error bars give the 1-$\sigma$ formal random fitting errors for each value. There is a ~30% uncertainty from the model parameters and calibration. The filled circles are from the observations of Jehin et al. (2022a,2023)

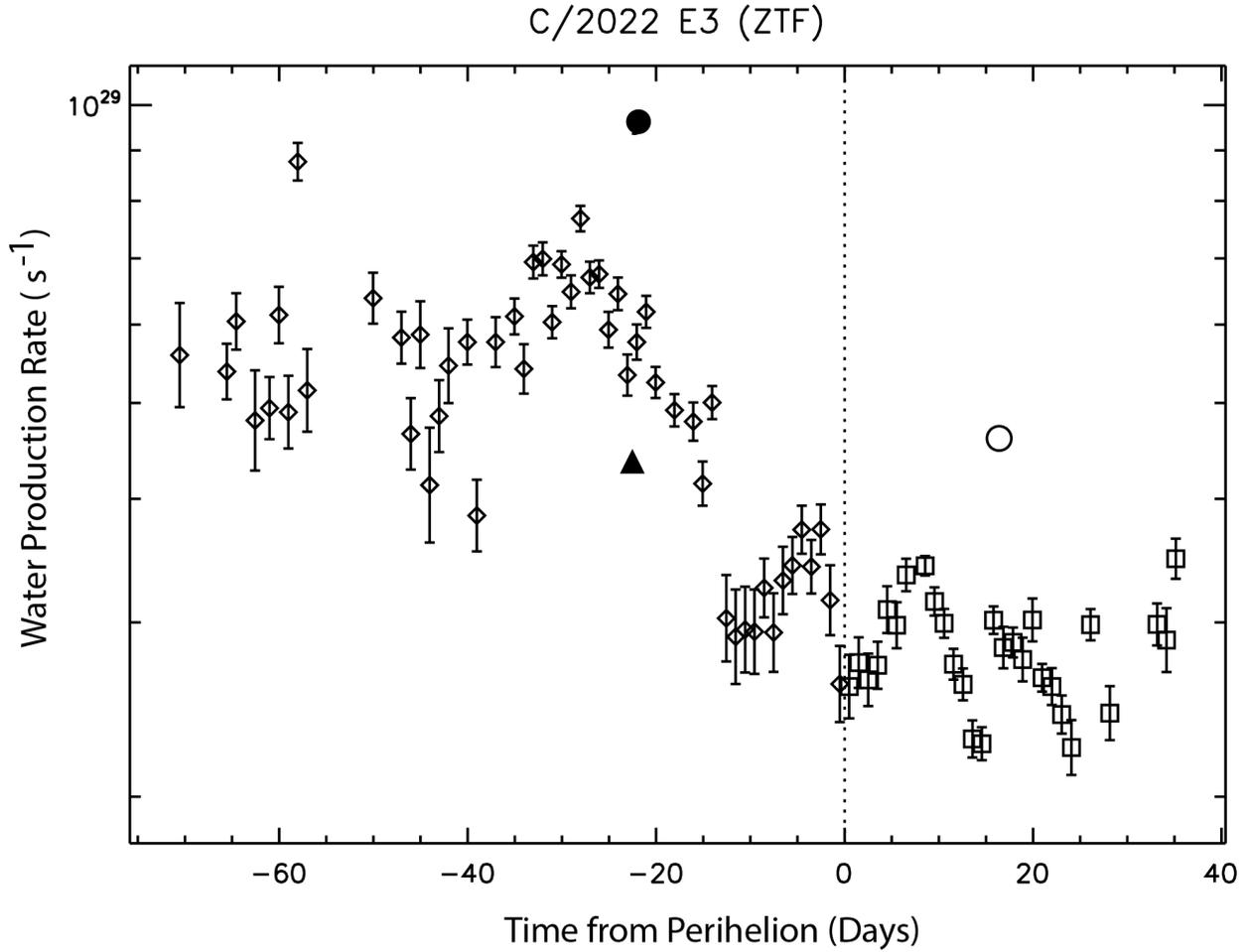

Figure 2. Water production rate in comet C/2023 E3(ZTF) as a function of time from perihelion. The open symbols points give the water production rate in s$^{-1}$ from single images. The error bars give the 1-$\sigma$ formal random fitting errors for each value. There is a ~30% uncertainty from the model parameters and calibration. The filled circle is from Biver et al. (2024), the filled triangle from Jehin et al. (2022b), and the open circle is from Schleicher (private communication).

Table 1

Summary of SOHO/SWAN Observations

| Comet | Perihelion | q(au) | Images | Start-End Date |
|---|---|---|---|---|
| C/2017 K2 (PanSTARRS) | 2022 Dec 19.69 | 1.797 | 83 | 28 Oct 2022-25 Apr 2023 |
| C/2022 E3 (ZTF) | 2023 Jan 12.78 | 1.112 | 80 | 2 Nov 2022-15 Feb 2023 |

Notes to Table 1

$r_H$ = heliocentric distance (au) range

q = perihelion distance (au)

Table 2

SOHO/SWAN Observations of C/2017 K2 (PanSTARRS) and Water Production Rates

| $\Delta T$ (Days) | R (au) | $\Delta$ (au) | g (s$^{-1}$) | Q ($10^{28}$ s$^{-1}$) | $\delta Q$ ($10^{28}$ s$^{-1}$) |
|---|---|---|---|---|---|
| -51.628 | 1.914 | 2.576 | 0.002190 | 7.49 | 0.07 |
| -49.628 | 1.906 | 2.581 | 0.002188 | 9.88 | 0.05 |
| -22.571 | 1.820 | 2.574 | 0.002133 | 9.17 | 0.06 |
| -20.571 | 1.816 | 2.568 | 0.002131 | 12.27 | 0.06 |
| -19.571 | 1.815 | 2.565 | 0.001522 | 13.05 | 0.05 |
| -18.571 | 1.813 | 2.562 | 0.002123 | 9.38 | 0.06 |
| 13.518 | 1.805 | 2.396 | 0.002082 | 14.06 | 0.01 |
| 14.518 | 1.807 | 2.390 | 0.002092 | 16.78 | 0.01 |
| 15.518 | 1.808 | 2.384 | 0.002080 | 15.02 | 0.01 |
| 16.518 | 1.810 | 2.378 | 0.002079 | 17.36 | 0.01 |
| 17.519 | 1.811 | 2.372 | 0.002065 | 13.12 | 0.01 |
| 18.519 | 1.813 | 2.366 | 0.002078 | 13.87 | 0.01 |
| 19.519 | 1.814 | 2.360 | 0.002067 | 11.34 | 0.02 |
| 20.519 | 1.816 | 2.354 | 0.002036 | 12.12 | 0.02 |
| 21.519 | 1.818 | 2.348 | 0.001838 | 13.59 | 0.02 |
| 22.546 | 1.820 | 2.342 | 0.002032 | 10.53 | 0.02 |
| 23.546 | 1.822 | 2.336 | 0.001997 | 12.69 | 0.01 |
| 24.546 | 1.824 | 2.330 | 0.001972 | 9.53 | 0.02 |
| 25.546 | 1.827 | 2.325 | 0.001881 | 16.47 | 0.01 |
| 26.546 | 1.829 | 2.319 | 0.001962 | 11.40 | 0.02 |
| 27.548 | 1.831 | 2.314 | 0.002021 | 16.29 | 0.01 |
| 28.548 | 1.834 | 2.308 | 0.002051 | 12.20 | 0.02 |
| 29.548 | 1.837 | 2.303 | 0.002084 | 14.02 | 0.01 |
| 30.548 | 1.839 | 2.297 | 0.002091 | 18.74 | 0.01 |
| 31.548 | 1.842 | 2.292 | 0.002104 | 13.29 | 0.01 |
| 33.575 | 1.848 | 2.282 | 0.002113 | 14.02 | 0.01 |
| 39.576 | 1.867 | 2.256 | 0.002242 | 11.58 | 0.01 |
| 40.577 | 1.871 | 2.253 | 0.002241 | 12.94 | 0.01 |
| 41.576 | 1.874 | 2.249 | 0.002227 | 12.01 | 0.01 |
| 42.576 | 1.878 | 2.246 | 0.002221 | 9.99 | 0.01 |
| 43.576 | 1.882 | 2.242 | 0.002199 | 9.48 | 0.01 |
| 44.576 | 1.885 | 2.239 | 0.001880 | 13.63 | 0.01 |
| 45.603 | 1.889 | 2.237 | 0.002094 | 9.54 | 0.01 |
| 46.604 | 1.893 | 2.234 | 0.002024 | 12.59 | 0.01 |
| 48.604 | 1.901 | 2.230 | 0.001830 | 9.43 | 0.02 |
| 49.604 | 1.906 | 2.228 | 0.001912 | 9.40 | 0.02 |
| 50.604 | 1.910 | 2.226 | 0.001969 | 11.36 | 0.01 |
| 52.605 | 1.918 | 2.224 | 0.002020 | 9.81 | 0.01 |
| 53.605 | 1.923 | 2.223 | 0.002061 | 9.29 | 0.01 |
| 54.605 | 1.927 | 2.222 | 0.002064 | 8.74 | 0.01 |
| 55.604 | 1.932 | 2.222 | 0.002098 | 7.13 | 0.02 |
| 56.605 | 1.937 | 2.222 | 0.002138 | 8.98 | 0.01 |
| 57.605 | 1.941 | 2.222 | 0.002161 | 7.79 | 0.01 |
| 58.604 | 1.946 | 2.222 | 0.002173 | 8.19 | 0.01 |
| 59.936 | 1.953 | 2.223 | 0.002170 | 7.26 | 0.01 |
| 60.936 | 1.958 | 2.225 | 0.002212 | 6.65 | 0.01 |
| 61.936 | 1.963 | 2.226 | 0.002259 | 6.92 | 0.01 |
| 62.936 | 1.968 | 2.228 | 0.002202 | 8.00 | 0.01 |
| 63.936 | 1.973 | 2.230 | 0.002283 | 11.22 | 0.01 |
| 64.936 | 1.978 | 2.232 | 0.002221 | 11.26 | 0.01 |
| 67.936 | 1.994 | 2.240 | 0.002124 | 14.85 | 0.01 |
| 68.936 | 1.999 | 2.244 | 0.002109 | 12.21 | 0.01 |
| 69.937 | 2.005 | 2.248 | 0.002050 | 10.46 | 0.01 |
| 70.937 | 2.011 | 2.252 | 0.002001 | 12.34 | 0.01 |
| 71.937 | 2.016 | 2.256 | 0.001995 | 7.91 | 0.02 |
| 72.937 | 2.022 | 2.261 | 0.001475 | 9.82 | 0.02 |

| ΔT | r | Δ | | | |
|---|---|---|---|---|---|
| 73.937 | 2.028 | 2.266 | 0.001475 | 6.89 | 0.03 |
| 74.937 | 2.033 | 2.271 | 0.001476 | 8.09 | 0.03 |
| 76.937 | 2.045 | 2.282 | 0.001476 | 8.50 | 0.03 |
| 77.937 | 2.051 | 2.288 | 0.001476 | 9.48 | 0.03 |
| 78.937 | 2.057 | 2.295 | 0.001476 | 10.94 | 0.02 |
| 79.937 | 2.063 | 2.301 | 0.001476 | 7.15 | 0.03 |
| 80.937 | 2.069 | 2.308 | 0.001477 | 10.16 | 0.02 |
| 81.937 | 2.075 | 2.315 | 0.001477 | 10.36 | 0.02 |
| 82.937 | 2.082 | 2.323 | 0.001477 | 9.76 | 0.03 |
| 83.937 | 2.088 | 2.331 | 0.001477 | 9.65 | 0.02 |
| 85.937 | 2.101 | 2.347 | 0.001477 | 6.66 | 0.04 |
| 86.937 | 2.107 | 2.356 | 0.001477 | 6.77 | 0.04 |
| 90.962 | 2.133 | 2.393 | 0.001478 | 9.42 | 0.02 |
| 101.196 | 2.203 | 2.504 | 0.001477 | 3.35 | 0.09 |
| 102.196 | 2.210 | 2.516 | 0.001477 | 5.32 | 0.06 |
| 103.196 | 2.217 | 2.529 | 0.001476 | 8.14 | 0.05 |
| 110.172 | 2.268 | 2.618 | 0.001473 | 5.19 | 0.08 |
| 111.168 | 2.275 | 2.632 | 0.001473 | 2.19 | 0.15 |
| 112.167 | 2.283 | 2.645 | 0.001473 | 3.74 | 0.11 |
| 117.168 | 2.320 | 2.715 | 0.001479 | 3.15 | 0.11 |
| 118.168 | 2.328 | 2.729 | 0.001479 | 3.07 | 0.11 |
| 119.144 | 2.335 | 2.743 | 0.001478 | 4.59 | 0.08 |
| 120.144 | 2.343 | 2.757 | 0.001478 | 3.29 | 0.11 |
| 123.144 | 2.366 | 2.801 | 0.001477 | 3.40 | 0.08 |
| 124.144 | 2.374 | 2.816 | 0.001476 | 7.51 | 0.05 |
| 126.297 | 2.391 | 2.847 | 0.001476 | 3.78 | 0.09 |
| 127.298 | 2.398 | 2.862 | 0.001475 | 4.80 | 0.07 |

Notes. ΔT (Days from Perihelion December 12.68, 2020)

r: Heliocentric distance (au)

Δ: Comet-SOHO distance (au)

g: Solar Lyman-α g-factor (photons s$^{-1}$) at 1 au

Q: Water production rates for each image (s$^{-1}$)

δQ: internal 1-sigma uncertainties

Table 3

SOHO/SWAN Observations of C/2022 E3 (ZTF) and Water Production Rates

| ΔT (Days) | r (au) | Δ (au) | g ($s^{-1}$) | Q ($10^{28}$ $s^{-1}$) | δQ ($10^{28}$ $s^{-1}$) |
|---|---|---|---|---|---|
| −70.561 | 1.572 | 2.123 | 0.001685 | 5.59 | 0.72 |
| −65.561 | 1.520 | 2.074 | 0.002243 | 5.38 | 0.36 |
| −64.560 | 1.510 | 2.064 | 0.002181 | 6.04 | 0.41 |
| −62.560 | 1.490 | 2.041 | 0.001665 | 4.80 | 0.59 |
| −61.029 | 1.475 | 2.023 | 0.002140 | 4.94 | 0.37 |
| −60.029 | 1.465 | 2.011 | 0.002100 | 6.13 | 0.42 |
| −59.028 | 1.455 | 1.998 | 0.002063 | 4.89 | 0.43 |
| −58.028 | 1.445 | 1.985 | 0.001648 | 8.77 | 0.39 |
| −57.028 | 1.436 | 1.972 | 0.001982 | 5.15 | 0.52 |
| −50.030 | 1.370 | 1.870 | 0.001939 | 6.38 | 0.39 |
| −47.030 | 1.344 | 1.821 | 0.001929 | 5.82 | 0.36 |
| −46.030 | 1.335 | 1.804 | 0.001889 | 4.65 | 0.40 |
| −45.031 | 1.326 | 1.786 | 0.001860 | 5.86 | 0.47 |
| −44.031 | 1.318 | 1.769 | 0.001833 | 4.13 | 0.59 |
| −43.031 | 1.310 | 1.750 | 0.001876 | 4.85 | 0.42 |
| −42.031 | 1.302 | 1.732 | 0.001898 | 5.45 | 0.50 |
| −40.031 | 1.286 | 1.694 | 0.002003 | 5.76 | 0.31 |
| −39.031 | 1.278 | 1.675 | 0.002045 | 3.85 | 0.33 |
| −37.031 | 1.262 | 1.635 | 0.002050 | 5.76 | 0.34 |
| −35.031 | 1.248 | 1.594 | 0.002087 | 6.12 | 0.26 |
| −34.031 | 1.241 | 1.573 | 0.002095 | 5.41 | 0.32 |
| −33.031 | 1.234 | 1.551 | 0.002099 | 6.94 | 0.27 |
| −32.031 | 1.227 | 1.529 | 0.002079 | 6.99 | 0.27 |
| −31.031 | 1.220 | 1.507 | 0.002084 | 6.03 | 0.23 |
| −30.032 | 1.214 | 1.485 | 0.002117 | 6.90 | 0.21 |
| −29.032 | 1.208 | 1.462 | 0.002089 | 6.47 | 0.25 |
| −28.032 | 1.201 | 1.439 | 0.002073 | 7.68 | 0.23 |
| −27.032 | 1.195 | 1.416 | 0.002033 | 6.70 | 0.25 |
| −26.055 | 1.190 | 1.393 | 0.002005 | 6.75 | 0.22 |
| −25.055 | 1.184 | 1.369 | 0.001985 | 5.93 | 0.25 |
| −24.055 | 1.179 | 1.345 | 0.001969 | 6.44 | 0.25 |
| −23.055 | 1.174 | 1.321 | 0.001966 | 5.34 | 0.26 |
| −22.055 | 1.169 | 1.296 | 0.001947 | 5.76 | 0.24 |
| −21.054 | 1.164 | 1.271 | 0.001927 | 6.18 | 0.23 |
| −20.054 | 1.159 | 1.246 | 0.001919 | 5.24 | 0.19 |
| −18.054 | 1.150 | 1.194 | 0.001957 | 4.92 | 0.19 |
| −16.061 | 1.143 | 1.142 | 0.002026 | 4.79 | 0.22 |
| −15.061 | 1.139 | 1.116 | 0.002001 | 4.14 | 0.22 |
| −14.061 | 1.136 | 1.089 | 0.002016 | 5.00 | 0.20 |
| −12.559 | 1.131 | 1.049 | 0.002079 | 3.03 | 0.32 |
| −11.558 | 1.128 | 1.021 | 0.002120 | 2.90 | 0.34 |
| −10.559 | 1.125 | 0.994 | 0.002155 | 2.95 | 0.31 |
| −9.558 | 1.123 | 0.966 | 0.002147 | 2.94 | 0.30 |
| −8.538 | 1.121 | 0.938 | 0.002159 | 3.25 | 0.23 |
| −7.539 | 1.119 | 0.910 | 0.002135 | 2.93 | 0.28 |
| −6.538 | 1.117 | 0.882 | 0.002090 | 3.31 | 0.27 |
| −5.539 | 1.116 | 0.854 | 0.002081 | 3.43 | 0.24 |
| −4.539 | 1.115 | 0.826 | 0.002093 | 3.72 | 0.21 |
| −3.528 | 1.114 | 0.798 | 0.002187 | 3.42 | 0.22 |
| −2.528 | 1.113 | 0.770 | 0.002216 | 3.73 | 0.22 |
| −1.528 | 1.113 | 0.741 | 0.002205 | 3.16 | 0.27 |
| −0.511 | 1.112 | 0.713 | 0.002206 | 2.60 | 0.24 |
| 0.489 | 1.112 | 0.685 | 0.002208 | 2.58 | 0.20 |
| 1.489 | 1.113 | 0.656 | 0.002233 | 2.73 | 0.17 |
| 2.501 | 1.113 | 0.628 | 0.002206 | 2.63 | 0.17 |
| 3.502 | 1.114 | 0.601 | 0.002163 | 2.72 | 0.15 |
| 4.517 | 1.115 | 0.573 | 0.001784 | 3.09 | 0.17 |

| ΔT | r | Δ | g | Q | δQ |
|---|---|---|---|---|---|
| 5.517 | 1.116 | 0.546 | 0.002095 | 2.98 | 0.16 |
| 6.531 | 1.117 | 0.519 | 0.001984 | 3.35 | 0.13 |
| 8.544 | 1.121 | 0.467 | 0.001948 | 3.42 | 0.08 |
| 9.544 | 1.123 | 0.442 | 0.001963 | 3.15 | 0.11 |
| 10.563 | 1.125 | 0.418 | 0.002001 | 2.99 | 0.10 |
| 11.563 | 1.128 | 0.395 | 0.002015 | 2.72 | 0.10 |
| 12.565 | 1.131 | 0.374 | 0.002057 | 2.60 | 0.10 |
| 13.566 | 1.134 | 0.355 | 0.002067 | 2.29 | 0.10 |
| 14.566 | 1.137 | 0.337 | 0.002090 | 2.26 | 0.09 |
| 15.811 | 1.142 | 0.319 | 0.002121 | 3.02 | 0.10 |
| 16.839 | 1.145 | 0.308 | 0.002129 | 2.83 | 0.14 |
| 17.866 | 1.150 | 0.299 | 0.002145 | 2.86 | 0.10 |
| 18.895 | 1.154 | 0.294 | 0.002201 | 2.75 | 0.14 |
| 19.929 | 1.158 | 0.294 | 0.002250 | 3.02 | 0.15 |
| 20.958 | 1.163 | 0.297 | 0.002287 | 2.64 | 0.09 |
| 21.987 | 1.168 | 0.304 | 0.002274 | 2.58 | 0.11 |
| 23.037 | 1.173 | 0.315 | 0.002282 | 2.42 | 0.11 |
| 24.066 | 1.179 | 0.328 | 0.002234 | 2.24 | 0.15 |
| 26.101 | 1.190 | 0.364 | 0.002166 | 2.98 | 0.11 |
| 28.146 | 1.202 | 0.407 | 0.002091 | 2.43 | 0.16 |
| 33.160 | 1.235 | 0.536 | 0.002066 | 2.99 | 0.15 |
| 34.158 | 1.242 | 0.563 | 0.002095 | 2.88 | 0.22 |
| 35.147 | 1.249 | 0.591 | 0.002094 | 3.48 | 0.17 |

Notes. ΔT (Days from Perihelion January 3.30, 2022)

r: Heliocentric distance (au)

Δ: Comet-SOHO distance (au)

g: Solar Lyman-α g-factor (photons s$^{-1}$) at 1 au

Q: Water production rates for each image (s$^{-1}$)

δQ: internal 1-sigma uncertainties

**Highlights**

SOHO/SWAN hydrogen Lyman-a camera observed comets C/2017 K2 (PanSTARRS) and C/2022 E3 (ZTF).

Water production rates were determined on 83 dates for comet K2 and 80 for comet E3.

Comet K2 varied with an exponent of -3.2 post-perihelion; the few pre-perihelion observations were similar.

Comet E3 varied irregularly with exponents of -1.3 pre- and -0.4 post-perihelion and 6 notable peaks.